\documentclass[secnumarabic,amssym,amsmath,amsfonts,nobibnotes,twoside,PRA]{revtex4}

\newcommand{\C}{\ensuremath{\mathbb{C}}}
\newcommand{\T}{\ensuremath{\mathbb{T}}}

\begin{document}

\title{Four-parameter families of complex Hadamard matrices of order six}
\author{P Di\c t\u a}
\email{dita@zeus.theory.nipne.ro}

\affiliation{National Institute of Physics and Nuclear Engineering,
P.O. Box MG6, Bucharest, Romania}

\begin{abstract}
In this paper we provide a general method to construct four-parameter families of  complex Hadamard  matrices of order six. Our approach is to write a 6-dimensional matrix as composed of four blocks, each one in the form of a circulant matrix.   Use of Sylvester  method of  inverse orthogonal matrices leads to two constraints on the six parameters, and their fulfilment generates complex Hadamard matrices     when the free  parameters  take values on the unit circle. We hope that the problem of mutually unbiased bases could be solved for six-level systems by using our results.
\end{abstract}

\maketitle

\section{Introduction}

Construction of  complex $6\times 6$ Hadamard  matrices has generated a considerable interest in the last years, this case being quite accessible,  leading to a few numerical matrices and to several matrices depending on few arbitrary phases, see Refs. \cite{A}-\cite{D},  \cite{Ha}, \cite{JMM}, \cite{K1}- \cite{MS},  \cite{Sz}-\cite{Sz1}, \cite{Z}. It was conjectured in paper  \cite{B} that the above results are particular cases of a more general four-parameter family yet to be discovered.  The number four is generated by the so called defect of a unitary matrix, see \cite{TZ}.

 It seems that until now there is only one attempt to construct such a family, see paper \cite{Sz1},  where his author starts with a dephased 6-dimensional matrix composed of four blocks and imposes some orthogonality conditions to get a Hadamard matrix. However no explicit form for such a matrix is provided.

The orthogonality concept is essential for Hadamard matrices and in the following  we will  use  the method of  inverse orthogonal matrices introduced  by   Sylvester in  paper \cite{JJS}  in the most general form. In this  paper   we will employ only  a   particular class of  inverse orthogonal matrices, $O=(o_{ij})$, namely   those matrices whose inverse  is given by 
\begin{eqnarray}O^{-1}=(1/o_{ij})^t=(1/o_{ji})\label{inv}\end{eqnarray}
 where $t$ means transpose, and their entries  $0\ne o_{ij}\in \C$ satisfy the relation
\begin{eqnarray} OO^{-1}= nI_n\label{inv1}\end{eqnarray}
 When the $ o_{ij} $  entries take values on the unit circle, $O^{-1} $ coincides with the Hermitean conjugate  $O^{*} $ of  $O $, and in this case relation (\ref{inv1}) is the definition of complex Hadamard matrices. Accordingly  the problem is to construct inverse orthogonal matrices that will provide the corresponding complex Hadamard matrices.

Our starting point is to write an arbitrary matrix as a four block matrix of the simplest form, as
\begin{eqnarray}
M=\left[ \begin{array}{cc}
A&B\\
C&D\end{array}\right]\label{Sy}\end{eqnarray}
similar to that from paper  \cite{JJS},
where $ A$ and $B$ are circulant matrices of arbitrary dimension $n$. As concerns the C and D matrices they are given by $C=B^{-1}$ and $D=- A^{-1}$, according to relation (\ref{inv}). The real problem is to find  those constraints on $ A$ and $B$ entries whose fulfilment transforms  $M$ into a complex Hadamard matrix, i.e. the new  $M$ matrix  has unimodular entries and satisfies the relation
\begin{eqnarray}
M M^{-1}= 2 n I_{2n} \label{inv2}
\end{eqnarray}

An important problem for complex  Hadamard matrices is that of their equivalence. Nowadays one makes use of two different equivalence methods: the standard method and the unitary equivalence.
The first one  is currently  applyed to real Hadamard matrices and has its origin in Sylvester paper  who introduced the so called standard form for real  matrices, whose  entries of the first row and column are equal to 1.  It was also extended to complex Hadamards. 

Its nowadays form says that two Hadamard matrices $H_1$ and $H_2$ are  equivalent, written as $H_1\equiv H_2$, if there exist two diagonal unitary matrices $D_1$ and $D_2$, and permutation matrices $P_1$ and $P_2$, such that 
\begin{eqnarray}
H_1=D_1P_1H_2P_2D_2\label{eq}\end{eqnarray}

However the complex Hadamard matrices naturally belong to the class of normal matrices. A matrix $N$ is normal if it commutes with its adjoint $N^*$, i.e. it satisfies the relation $N\,N^*= N^*\,N$. For this class of operators the unitary equivalence takes a simple form and says that every normal matrix is similar to a diagonal matrix $D$, which means that there exists a unitary matrix $U$, such that
\begin{eqnarray} N= UDU^* \label{nor}\end{eqnarray}
 see \cite{P} p. 357, or \cite{K}. 

Two important classes of normal operators are the unitary and self-adjoint matrices, such that for both these classes two matrices $M_1$ and $M_2$  are unitary equivalent
 iff they have the same spectrum, or equivalently, the characteristic polynomials are the same up to a multiplicative constant factor, see  \cite{P},  \cite{K}, i.e.
\begin{eqnarray}
det(x\,I_n-M_1/\sqrt{n})=det(x\,I_n-M_2/\sqrt{n})\label{unitar}\end{eqnarray}
where $det$ is the determinant of the corresponding matrix.

For unitary matrices the entries of the diagonal matrix $D$ are unimodular, and for self-adjoint they are real numbers.

 To understand better the difference between the two equivalence methods  we give a simple example of a selfadjoint Hadamard matrix whose entries from the first row and column are equal to 1.
\begin{eqnarray}
d_6=\left[ \begin{array}{rrrrrr}
1&1&1&1&1&1\\
1&-1&i&i&-i&-i\\
1&-i&-1&1&-1&i\\
1&-i&1&-1&i&-1\\
1&i&-1&-i&1&-1\\
1&i&-i&-1&-1&1
\end{array}\right]\label{no}\end{eqnarray}
The unitary equivalence implies the knowledge of its spectrum, which in this case is given by
\begin{eqnarray} Sp(d_6)=\left[-1^3,1^3\right]\label{spm}\end{eqnarray}
where power means eigenvalue multiplicity. The above result shows that $d_6$ is a complex Hadamard matrix, and in the same time a selfadoint matrix, its spectrum being unimodular and real.

 If we make use  of the standard equivalence (\ref{eq})  the following matrix 
\begin{eqnarray}
d_{61}=\left[ \begin{array}{rrrrrr}
1&1&1&1&1&1\\
1&-1&1&-1&i&-i\\
1&1&-1&i&-1&-i\\
1&-i&-1&-1&1&i\\
1&-1&-i&1&-1&i\\
1&i&i&-i&-i&-1
\end{array}\right]\label{nor1}\end{eqnarray}
 is equivalent to (\ref{no}), but its spectrum  is given by
\begin{eqnarray} Sp(d_{61})=\left[-1^2,1^2,\frac{i -\sqrt{2}}{\sqrt{3}} ,-\frac{i+\sqrt{2}}{\sqrt{3}}\right]\label{spm1}\end{eqnarray}
Thus when  one makes use of the usual equivalence, see  relation (\ref{eq}), the classical quantum mechanics will be into a big danger, because the $d_{61}$ eigenvalues are unimodular, but not real.

In the first case $d_{6} = d_{6}^*$, and in the second case  $d_{61} \ne d_{61}^*$.  Thus the change of rows and/or colums between themselves modifies the matrix symmetry, and accordingly its spectrum.

The paper is organized as follows. In Sec. {\bf 2} we treat the simplest case, namely that when $A$ and $B$ matrices are 2-dimensional circulant matrices to see how the method  works. In this simple case  we show that the Sylvester orthogonality is equivalent  to one condition  on the four arbitrary parameters entering $A$ and $B$ matrices which leads to four non dephased and  inequivalent complex Hadamard matrices.  Similar results for  $n=6$ are given  in  Sec.  {\bf 3}.  In  Sec.  {\bf 4} we show how the results obtained in sections  {\bf 2} and  {\bf 3} lead to higher order Hadamard matrices. The paper ends by a few conclusions.

\section{Four-dimensional Hadamard matrices}

The current opinion  about four-dimensional Hadamard matrices is that up to equivalence there is only one matrix, namely that found by Hadamard \cite{Had},  which is written in a dephased form. As we said this approach has its source in Sylvester paper \cite{JJS} who introduced the so called standard form for real $\pm 1$ matrices.  This equivalence was essentially extended  to all complex Hadamard matrices.
 But if we look at Hadamard matrices as unitary matrices, then their equivalence is given by their spectra,  according to papers \cite{P}, or \cite{K},  and the problem changes significantly.

If one takes $M$ of the  form
\begin{eqnarray}
M_4=\left[ \begin{array}{rrrr}
a&b&c&d\\*[2mm]
-b&a&-d&c\\*[2mm]
\frac{1}{c}&-\frac{1}{d}&-\frac{1}{a}&\frac{1}{b}\\*[2mm]
\frac{1}{d}&\frac{1}{c}&-\frac{1}{b}&-\frac{1}{a}\end{array}\right]\label{Di}\end{eqnarray}
 the relation (\ref{inv2}) leads to the following orthogonality condition 
\begin{eqnarray}
(b c+ a d)(a c-b d)=0\label{ort}\end{eqnarray}
and if it is satisfied, one gets an orthogonal matrix. As it is easily seen the above equation has eight solutions, e.g. $a=\frac{bd}{c}$, $ b=-\frac{ad}{c}$, $c=\frac{bd}{a}$, etc,  but the unitary equivalence principle restricts them to four nonequivalent  ones.

 If we start with
$a=\frac{bd}{c}$ the orthogonal matrix has the form
\begin{eqnarray}
H_{4}(b,c,d)=\left[\begin{array}{rrrr}
\frac{b d}{c}&b&c&d\\*[2mm]
-b&\frac{b d}{c}&-d&c\\*[2mm]
\frac{1}{c}&-\frac{1}{d}&-\frac{c}{b d}&\frac{1}{b}\\*[2mm]
\frac{1}{d}&\frac{1}{c}&-\frac{1}{b}&-\frac{c}{b d}\end{array}\right]\label{h41}\end{eqnarray}
When the parameters $ b,\,c,\,d \in \T^3$, the three dimensional torus, the  matrix (\ref{h41}) is complex Hadamard  and its spectrum  is given by the solutions of the equation
\begin{eqnarray}
ec_1(b,c,d;x)& =& x^4-\frac{ (b^2d^2-c^2)}{bcd} x^3+\frac{(c^2+d^2)(c^2+b^4d^2)-8b^2c^2d^2}{4b^2c^2d^2}x^2
+\frac{ (b^2d^2-c^2)}{bcd}  x + 1 =0\label{ec1}
\end{eqnarray}
which depends on three independent phases. The above equation can be easily solved. However since (\ref{ec1}) is a reciprocal equation by the substitution $1/x-x=y$ it is reduced to a second order equation. This property is almost generic  for any complex Hadamard matrix  obtained by using relation (\ref{inv2}), such that its spectrum can be explicitly computed for dimensions $n=4,\,6,\,8$. However there are few cases for dimensions $n\ge 6 $ when only the numeric spectrum can be obtained, and they will be discussed in an other paper.

Indeed the above substitution transforms (\ref{ec1})  into the following simpler equation

\begin{eqnarray}
ec_1(b,c,d;x) = y^2-\frac{ (b^2d^2-c^2)}{bcd}y+\frac{(c^2+d^2)(c^2+b^4d^2)}{4b^2c^2d^2} =0\label{ec11}
\end{eqnarray}
 and its solutions are
\begin{eqnarray}
y_{1,2}=\frac{(b^2d^2-c^2)\pm \sqrt{-(1+b^2)^2c^2d^2}}{2\,b\,c\,d}
\end{eqnarray}
which provide the spectrum of equation (\ref{ec1})
\begin{eqnarray}
x_{1,2}&=&\frac{b^2d^2-c^2-\sqrt{-(1+b^2)^2c^2d^2}\pm \sqrt{16\,b^2c^2d^2+\left(c^2-b^2d^2+\sqrt{-(1+b^2)^2c^2d^2}\right)^2}}{4\,b\,c\,d},\\
 x_{3,4}&=&\frac{b^2d^2-c^2-\sqrt{-(1+b^2)^2c^2d^2}\pm \sqrt{16\,b^2c^2d^2+\left(c^2-b^2d^2-\sqrt{-(1+b^2)^2c^2d^2}\right)^2}}{4\,b\,c\,d}  \end{eqnarray}

For example the parameters   $b=1,\,c=i,\,d=-i$ provide the spectrum
 \begin{eqnarray}
Sp(H_{4}(1,i,-i))=\left[x_{1,2}=-\frac{i \pm \sqrt{3}}{2},~~x_{3,4}=\frac{i \pm \sqrt{3}}{2}\right]\end{eqnarray}
where in the following $i=\sqrt{-1}$.

Equation (\ref{ec1}) for  $b=\pm 1,\,c= e^{\frac{i\pi}{2}},\,d = e^{\frac{i \pi}{4}}$ transforms into 
\begin{eqnarray}
2x^4+2 i\sqrt{2}x^3-3 x^2-2 i\sqrt{2}x+2=0\end{eqnarray}
and its spectrum is given by
\begin{eqnarray}
Sp(H_{4}(1,e^{\frac{i\pi}{2}},e^{\frac{i \pi}{4}}))=\left[\frac{(2-\sqrt{2})i\pm \sqrt{2(5+2 \sqrt{2}})}{4},\frac{-(2+\sqrt{2})i\pm \sqrt{2(5-2 \sqrt{2}})}{4} \right]\end{eqnarray}
The standard form  of $H_{4}$ matrix is
\begin{eqnarray}
H_{4a}=\left[ \begin{array}{rrrr}
1&1&1&1\\*[2mm]
1&-q&q&-1\\*[2mm]
1&-1&-1&1\\*[2mm]
1&q&-q&-1\end{array}\right]\label{ha}\end{eqnarray}
where $q=d^2/c^2$ and its spectrum is given by
\begin{eqnarray}
Sp(H_{4a}(q))=\left[-1,1,-\frac{1+q+\sqrt{1-14 q+q^2}}{4},-\frac{1+q-\sqrt{1-14 q+q^2}}{4}\right]\label{ha1}\end{eqnarray}
with $q$  an arbitrary phase. From the above formula one finds
\begin{eqnarray}
Sp(H_{4a}(1))~~&=&\left[-1,1,-\frac{1+ i\sqrt{3}}{2},\frac{-1+ i\sqrt{3}}{2}\right]\nonumber\\
Sp(H_{4a}(i))~~&=&\left[-1,1,-\frac{1+ i}{4}\pm\frac{(1-i)\sqrt{7}}{4}\right]\label{mult1}\\
Sp(H_{4a}(-1))&=&\left[-1^2,1^2 \right]\nonumber
\end{eqnarray}
 The spectrum of the original Hadamard matrix found in paper \cite{Ha} is obtained from formula (\ref{ha1}) by the change $q\rightarrow -q$.
If one interchange the third and the fourth rows of (\ref{ha}) between themselves the spectrum is given by
$[-1,1,1,-q]$. This example shows that by bringing a unitary matrix to its dephased  form, or by permuting rows and columns between themselves, the obtained matrices are in general not unitary equivalent with the  matrix one started with.

The parameter choices, $c=\frac{bd}{a}$, $d=\frac{ac}{b}$, $c=-\frac{a d}{b}$, and $ d=-\frac{b c}{a} $,   lead to only one matrix, $H_{42}(a,b)$, which depends on two parameters, and its  spectral  equation is
\begin{eqnarray}
ec_2(a,b;x) =x^4-\frac{a^2-1}{a}x^3+\frac{(a^2+b^2)(1+a^2b^2)-8a^2b^2}{4 a^2 b^2}x^2+\frac{a^2-1}{a} x+1=0\label{ec2}\end{eqnarray}

The matrix  obtained with the condition $b=\frac{ac}{d}$, $H_{43}(a,c,d)$,  has the  spectrum given by the equation
\begin{eqnarray}
ec_3(a,c,d;x)& =&   x^4 -\frac{a^2-1}{a}x^3 + \frac{(c^2+d^2)(a^4c^2+d^2)-8a^2c^2d^2}{4a^2c^2d^2}x^2
+\frac{a^2-1}{a} x + 1 = 0\label{ec3}\end{eqnarray}

The choice $a=-\frac{bc}{d}$ leads to $H_{44}(a,c,d)$ matrix with the spectral equation 
\begin{eqnarray}
ec_4(b,c,d;x)& = &x^4 +\frac{b^2c^2 - d^2}{b c d}x^3+\frac{(c^2 + d^2)(b^4 c^2 + d^2)-8b^2c^2d^2}{4 b^2c^2d^2} x^2
-\frac{b^2c^2 - d^2}{b c d} x + 1=0
\label{ec4}\end{eqnarray}

The choice  $b=-\frac{a d}{c}$ provides the matrix $H_{45}(a,c,d)$ whose  spectral equation is 

\begin{eqnarray}
ec_5(b,c,d;x)& = &x^4- \frac{(a^2-1)}{a}x^3+
\frac{(c^2 + d^2)( c^2 +a^4 d^2)-8b^2c^2d^2}{4 a^2c^2d^2} x^2
+\frac{(a^2-1)}{a}x + 1=0
\label{ec5}\end{eqnarray}
equation which is similar to equation (\ref{ec3}).

 A more careful analysis shows that the spectra of equations (\ref{ec2}) and (\ref{ec3}) coincide on the surface $d^2 = a^2b^2c^2$, i.e. the matrix $H_{42}(a,b)$  is a particular case of the matrix  $H_{43}(a,c,d)$.

\section{Six-dimensional Hadamard matrices}

In order to see a few subtleties of case $n =  6$ we start with the Bj\"orck and Fr\"oberg  6-dimensional matrix, \cite{BF}, written in the form provided by Haagerup, \cite{Ha}
\begin{eqnarray}BF=
\left[\begin{array}{rrrrrr}
1&\frac{i}{d}&-\frac{1}{d}&-i&- d&i d\\*[2mm]
i d& 1&\frac{i}{d}&-\frac{1}{d}&-i&- d\\*[2mm]
- d&i d& 1&\frac{i}{d}&-\frac{1}{d}&-i\\*[2mm]
-i &- d&i d& 1&\frac{i}{d}&\-\frac{1}{d}\\*[2mm]
-\frac{1}{d}&-i &- d&i d& 1&\frac{i}{d}\\*[2mm]
\frac{i}{d}& -\frac{1}{d}&-i &- d&i d& 1 \end{array}\right]\label{BF}
\end{eqnarray}
The orthogonality relation (\ref{inv2}) leads to {\em one} constraint
\begin{eqnarray}
d^4-2 d^3-2 d+1=0\end{eqnarray}
whose solutions are 
\begin{eqnarray}
\left[d_{1,2}=\frac{1\pm i\sqrt{2}~3^{1/4}- \sqrt{3}} {2},\,\,d_{3,4}=\frac{1\pm \sqrt{2}~ 3^{1/4}+ \sqrt{3}}{2} \right]\end{eqnarray}
 The first two solutions are complex and unimodular and lead to {\em one} complex Hadamard matrix. The  $d_{3,4}$ solutions being real numbers different from $\pm 1$ lead to two different orthogonal matrices.
The spectral equation is the same for both $d_{1,2}$, and its form is
\begin{eqnarray}
x^6 -\sqrt{6}\,x^5+3\,x^4-2 \sqrt{2}x^3+3 \,x^2 -\sqrt{6}\,x+1=0 \label{ee1}
\end{eqnarray}
and the eigenvalues are  given by

\begin{eqnarray}Sp(BF)=\left[\frac{1\pm i}{\sqrt{2}},\frac{\sqrt{6}-\sqrt{2}+2\,3^{\frac{1}{4}}}{4}\pm i \sqrt{\frac{1-\sqrt{2\,\sqrt{3}-3}}{2}},\pm\frac{\sqrt{6}-\sqrt{2}+2\,3^{\frac{1}{4}}}{4}i +\sqrt{\frac{1-\sqrt{2\,\sqrt{3}-3}}{2}}\right]\end{eqnarray}

The entries of the two Sylvester orthogonal matrices are not unimodular so they do not generate Hadamard matrices.

The dephased form of the matrix (\ref{BF}) is 
\begin{eqnarray}bf=
\left[\begin{array}{rrrrrr}
1&1&1&1&1&1\\*[2mm]
1&-1&-\frac{1}{d}&-\frac{1}{d^2}&\frac{1}{d^2}&\frac{1}{d}\\*[2mm]
1&-d&1&\frac{1}{d^2}&-\frac{1}{d^3}&\frac{1}{d^2}\\*[2mm]
1&-d^2&d^2&-1&\frac{1}{d^2}&-\frac{1}{d^2}\\*[2mm]
1&d^2&-d^3&d^2&1&-\frac{1}{d}\\*[2mm]
1&d&d^2&-d^2&-d&-1 \end{array}\right]\label{bf}
\end{eqnarray}
and if we substitute in it, e.g. the $d_1-$root, its spectrum is $Sp(bf)=\left[-1^3, 1^3\right]$. Thus the matrices (\ref{BF}) and  (\ref{bf}) are not unitary equivalent.

The  6-dimensional matrix is of the form (\ref{Sy}) whith  $A$ and $B$ blocks 3-dimensional circulant matrices, and explicitly it is written as

\begin{eqnarray}
M_6=\left[\begin{array}{rrrrrr}
a&b&c&d&e&f\\*[2mm]
c&a&b&f&d&e\\*[2mm]
b&c&a&e&f&d\\*[2mm]
\frac{1}{d}&\frac{1}{f}&\frac{1}{e}&-\frac{1}{a}&-\frac{1}{c}&-\frac{1}{b}\\*[2mm]
\frac{1}{e}&\frac{1}{d}&\frac{1}{f}&-\frac{1}{b}&-\frac{1}{a}&-\frac{1}{c}\\*[2mm]
\frac{1}{f}&\frac{1}{e}&\frac{1}{d}&-\frac{1}{c}&-\frac{1}{b}&-\frac{1}{a}
\end{array}\right]\label{m6}\end{eqnarray}
 The orthogonality relation (\ref{inv2}) leads to two constraints
\begin{eqnarray}
 a\, b\, c\, d^2\, e+a^2 \,b\, d\, e\, f+b^2\, c\, d\, e\, f+a\, c^2\, d\, e\, f+a\, b\, c\, e^2\, f+a\, b\, c\, d\, f^2=0\label{con}\\
a\, b\, c\, d \,e^2+a\, b\, c\, d^2\, f+a\, b^2\, d\, e\, f+a^2\, c\, d\, e\, f+b\, c^2\,d\, e\, f+ a\, b\, c\, e\, f^2=0 \label{con1}
\end{eqnarray}
and when they are satisfied the matrix (\ref{m6}) gets orthogonal. Each equation is quadratic in $ a,\,b,\,c,\,d,\,e,\,f\,$ parameters such that one could solve one of them with respect to a chosen parameter. For example we can use   equation (\ref{con}) to solve the constraint  on  $f$ parameter, that provides two solutions for it
\begin{eqnarray}
f_{\pm}=\frac{-e(a^2bd -b^2cd-ac^2d-abce) \pm\sqrt{-4a^2b^2c^2d^3e+e^2(a^2bd+b^2cd+ac^2e+abce)^2}}{2 a b c d}\label{ecfpm}\end{eqnarray}
where the subscript of $f_{\pm}$ is related to the $\pm$ sign which appears in front of the square root in relation (\ref{ecfpm}). When both solutions $f_{\pm}$ are introduced in the equation (\ref{con1}) one gets only {\it one}  condition
\begin{eqnarray}
-a\,b\,c\,d^3-a\,b^2\,d^2\,e-a^2\,c\,d^2\,e-b\,c^2\,d^2\,e+a^2\,b\,d\,e^2+b^2\,c\,d\,e^2+a\,c^2\,d\,e^2+a\,b\,c\,e^3=0 \label{con2}\end{eqnarray}

If one solves equation (\ref{con1}) with respect to  $f$ parameter, and if these solutions are introduced in relation  (\ref{con}) one gets again the relation  (\ref{con2}).

This equation has a quadratic dependence in $a,\,b\,{\rm and}\,c$ parameters, and a cubic dependence in $d\,{\rm and}\, e$ parameters. Thus we can  solve it with respect to any of the first three parameters by getting six solutions for $a_{\pm}$, $b_{\pm}$, and $c_{\pm}$, and six solutions, $d_{1,2,3}$ and  $e_{1,2,3}$, for cubic parameters. For example  $a_{\pm}$ have the form
\begin{eqnarray}
a_{\pm}=\frac{(cd+be)(bd^2-ce^2)\pm\sqrt{(cd+be)^2(bd^2-ce^2)^2-4abcd^2e^2(cd-be)}}{2de(be-cd)}\label{ecapm}
\end{eqnarray}

If we solve equation  (\ref{con}) with respect to parameter $a$  one finds
\begin{eqnarray}
a_{\pm}=\frac{-c(b\,d^2\,e+c\,d\,e\,f+b\,e^2\,f+b\,d\,f^2) \pm\sqrt{-4\,b^3c\,d^2\,e^2\,f^2 + c^2(b\,d^2\,e+c\,d\,e\,f+b\,e^2\,f+b\,d\,f^2 )}}{2 a b c d}\label{ecapm1}\end{eqnarray}
When $a_{+}$ is introduced into  equation (\ref{con1}) one gets the following condition
\begin{eqnarray}
-b\,c^2\,d^2\,e +b^2\,c\,d\,e^2+b^2\,c\,d^2\,f+b^3\,d\,e\,f-c^3\,d\,e\,f-b\,c^2\,e^2\,f-b\,c^2\,d\,f^2+b^2\,c\,e\,f^2=0\label{ecfpm2}\end{eqnarray}
and if we solve it for $f$ parameter, its solutions are
\begin{eqnarray}
f_{\pm}=\frac{(cd+be)(bd^2-ce^2)\pm\sqrt{(cd+be)^2(bd^2-ce^2)^2-4abcd^2e^2(cd-be)}}{2de(be-cd)}\label{ecfpm3}
\end{eqnarray}
and two solutions for $d_{\pm}$ and  $e_{\pm}$, and three solutions for $b_{1,2,3}$ and  $c_{1,2,3}$.

The $a_{\pm}$ solutions from the first case, (\ref{ecapm}), coincide with the $f_{\pm}$ solutions from the second case, (\ref{ecfpm3}),   whilst the solutions for  $f_{\pm}$ from the first case, (\ref{ecfpm}), are different from the solutions for $a_{\pm}$, (\ref{ecapm1}). Thus the above coincidence suggests that there are only four independent matrices instead of eight for each pair $(a,f)$, $(d,e)$, etc.  To prove it one has to find  the spectral equation for each case, which is not a simple problem. 

 In contradistinction with the precedent four dimensional case the obtained matrices are large having a complicated dependence on $a_{\pm}$ and  $f_{\pm}$, as it is seen from relations (\ref{ecfpm}),   (\ref{ecapm}), (\ref{ecapm1}), (\ref{ecfpm3}), such that they cannot be written on an usual paper sheet.

  To have an idea about their spectra we make use of numerical values for the remaining parameters. 
For example if one chooses for  free parameters the following values, $b = 1,\,c = i,\,d = e^{\frac{\pi i}{4}},e =-1$, one get four numerical Hadamard matrices, and their spectra are given by the equations
\begin{eqnarray}
ec(f_{+},a_{+},x)& =& x^6+i \sqrt{3}\,x^5-3\,x^4-\frac{1}{3}i \sqrt{86+\frac{32\sqrt{2}}{3}}\,x^3+3\,x^2+i\sqrt{3}\,x-1=0\label{fpap}\\
ec(f_{+},a_{-},x)& =& x^6-i \sqrt{3}\,x^5-3\,x^4+\frac{1}{3}i \sqrt{86-\frac{32\sqrt{2}}{3}}\,x^3+3\,x^2-i\sqrt{3}\,x-1=0\label{fpam}\\
ec(f_{-},a_{+},x)& =& x^6+i \sqrt{3}\,x^5-3\,x^4-\frac{1}{3}i \sqrt{86+\frac{32\sqrt{2}}{3}}\,x^3+3\,x^2+i\sqrt{3}\,x-1=0\label{fmap}\\
ec(f_{-},a_{-},x)& =& x^6-i \sqrt{3}\,x^5-3\,x^4+\frac{1}{3}i \sqrt{86-\frac{32\sqrt{2}}{3}}x^3+3\,x^2-i\sqrt{3}\,x-1=0\label{fmam}\end{eqnarray}
where $f_{\pm}$ and $a_{\pm}$  are provided by relations (\ref{ecfpm}), and respectively (\ref{ecapm}). By the substitution $y=\frac{1}{x}-x$ the above equations transform into reduced equations
\begin{eqnarray}
y^3- i \sqrt{3} y^2+i\frac{-18+\sqrt{2(129+16\sqrt{2})}}{\sqrt{3\sqrt{3}}}=0,\,~~
y^3+ i \sqrt{3} y^2-i \frac{-18+2\sqrt{2(129-16\sqrt{2})}}{3\sqrt{3}}=0\label{fa}
\end{eqnarray}

If one makes use of the second set $(a_{\pm},f_{\pm})$ one gets

\begin{eqnarray}
 ec(a_{+},f_{+},x)& =& x^6+i \sqrt{3}x^5-3 x^4-\frac{i}{3}\sqrt{86+\frac{32\sqrt{2}}{3}}x^3+3x^2+i\sqrt{3}x -1=0\label{apfp}\\
 ec(a_{+},f_{-},x)& =& x^6-i \sqrt{3}x^5-3 x^4+\frac{i}{3}\sqrt{86-\frac{32\sqrt{2}}{3}}x^3+3x^2-i\sqrt{3}x -1=0\label{apfm}\\
 ec(a_{-},f_{+},x)& =& x^6-i \sqrt{3}x^5-3 x^4+\frac{i}{3}\sqrt{86-\frac{32\sqrt{2}}{3}}x^3+3x^2-i\sqrt{3}x -1=0\label{amfp}\\
 ec(a_{-},f_{-},x)& =& x^6+i \sqrt{3}x^5-3 x^4-\frac{i}{3}\sqrt{86+\frac{32\sqrt{2}}{3}}x^3+3x^2+i\sqrt{3}x -1=0\label{amfm}
\end{eqnarray}
and the reduced equations are the same as in the preceding case.

The above numerical examples could suggest that the pairs  $(f_{+},a_{\pm})$ and  $(a_{-},f_{\pm})$  lead to different spectra.
Taking into account that the relations (\ref{con}) and (\ref{con1}) can be solved for all pairs of parameters it follows that, in general, it could be  $4\times C_6^2 =4\times 15=60$ inequivalent matrices. 

 Supplementary solutions can be obtained from equations similar to (\ref{con2}) by solving them with  respect to cubic parameters; in the case of equation (\ref{con2}), the $d$ or $e$ parameters. In all this  cases the solutions are more complicated, and the corresponding matrices also. For example if we solve equation  (\ref{con2}) with respect to {\em e} parameter one expects to find three solutions. However in some numerical cases the  spectral equation does not depend on $f_{\pm}$, or $e_{1,2,3}$ solutions of the cubic equation.   For example  the  parameters, $a = 1,\, b = e^{\frac{2 \pi i}{3}},\,c = e^{\frac{4 \pi i}{3}},\,d = 1$, lead to the spectral equation whose spectrum is simple
\begin{eqnarray}
ec(f_{\pm},e_{1,2,3})= (x^2-1)(x^4+1)=0,\,\,Sp(f_{\pm},e_{1,2,3})=\left[-1,\,1,\,\pm \frac{1+i}{\sqrt{2}},\,\pm \frac{1-i}{\sqrt{2}} \right]\label{uq1}\end{eqnarray}

If we solve equation (\ref{con2}) with respect to $d$ parameter on finds for $a = 1,\, b = e^{\frac{2 \pi i}{3}},\,c = e^{\frac{4 \pi i}{3}},\,e = 1$  the following spectrum 
\begin{eqnarray}Sp(f_{\pm},d_{1,2,3})=\left[-1^3,\,1^3\right]\label{bf1}\end{eqnarray}
which coincides with the spectrum of the matrix (\ref{bf}). Thus the dephased form of  Bj\"orck and Fr\"oberg  matrix is unitary equivalent to a particular case of the matrix (\ref{m6}) when a second parameter is obtained from a cubic equation.
 Thus for each parameter, $a,\,b,\,c,\,d,\,e,\,f$, obtained from the solutions of  cubic equations, one gets at least a single Hadamard matrix, such that the number of all nonequivalent matrices could be at least $60+6=66$.

The above matrices are a rich source  for three-parameter Hadamard matrices. The simplest case will be to equal two parameters; however the obtained  matrix is also large. Many simpler matrices which depend on three parameters do exist and in the following  we give a few examples.

Looking at relation   (\ref{ecapm}) one easily sees  that if the following equation is satisfied
\begin{eqnarray}(b\,e+c\,d)(b\,d^2-c\,e^2)=0\label{hu} \end{eqnarray}
we can obtain three-parameter Hadamard matrices as follows. With the solution $ b= -cd/e$ from the first factor (\ref{hu}) one gets the Hadamard matrix
\begin{eqnarray}
D_{61}(c,d,e)=\left[\begin{array}{cccccc}
-\frac{\sqrt{c^4\,d^5\,e}}{c\,d^2\,e}&-\frac{c\,d}{e}&c&d&e&\frac{e^2\sqrt{-\frac{c^6\,d^6}{e^2}}}{c\sqrt{c^4\,d^5\,e}}\\*[2mm]
c&-\frac{\sqrt{c^4\,d^5\,e}}{c\,d^2\,e}&-\frac{c\,d}{e}&\frac{e^2\sqrt{-\frac{c^6\,d^6}{e^2}}}{c\sqrt{c^4\,d^5\,e}} &d&e\\*[2mm]
-\frac{c\,d}{e}&c&-\frac{\sqrt{c^4\,d^5\,e}}{c\,d^2\,e}&e&\frac{e^2\sqrt{-\frac{c^6\,d^6}{e^2}}}{c\sqrt{c^4\,d^5\,e}} &d\\*[2mm]
\frac{1}{d}&\frac{c\sqrt{c^4\,d^5\,e}}{e^2\sqrt{-\frac{c^6\,d^6}{e^2}}}&\frac{1}{e}&\frac{c\,d^2\,e}{\sqrt{c^4\,d^5\,e}}&-\frac{1}{c}&\frac{e}{c\,d}\\*[2mm]
\frac{1}{e}&\frac{1}{d}& \frac{c\sqrt{c^4\,d^5\,e}}{e^2\sqrt{-\frac{c^6\,d^6}{e^2}}}  &\frac{e}{c\,d}&\frac{c\,d^2\,e}{\sqrt{c^4\,d^5\,e}}&-\frac{1}{c}\\*[2mm]
 \frac{c\sqrt{c^4\,d^5\,e}}{e^2\sqrt{-\frac{c^6\,d^6}{e^2}}}  &\frac{1}{e}&\frac{1}{d}&-\frac{1}{c}&\frac{e}{c\,d}&\frac{c\,d^2\,e}{\sqrt{c^4\,d^5\,e}}
\end{array}\right]\label{mcde}\end{eqnarray}
which comes from the pair $(f_{+},\,a_{+})$  in this order. Its reduced polynomial has the form
\begin{eqnarray}
pol(f_{+},\,a_{+})= y^3-\frac{\sqrt{3/2}(c^2d-e)\sqrt{c^4d^5e}}{c^3\,d^3\,e} y^2+\frac{(c^4d^2+e^2)y}{c^2\,d\,e} -\nonumber\\ 
\frac{c^2d(d^3-e^3)(c^2d+e)^3+4(c^6d^3-e^3)\sqrt{c^4d^5e}}{6\,\sqrt{6\,}c^5\,d^4\,e^3}~~~\label{pol1}\end{eqnarray}

The next matrix comes from the pair  $(f_{+},\,a_{-})$ 
\begin{eqnarray}
D_{62}(c,d,e)=\left[\begin{array}{cccccc}
\frac{\sqrt{c^4\,d^5\,e}}{c\,d^2\,e}&-\frac{c\,d}{e}&c&d&e&-\frac{e^2\sqrt{-\frac{c^6\,d^6}{e^2}}}{c\sqrt{c^4\,d^5\,e}}\\*[2mm]
c&\frac{\sqrt{c^4\,d^5\,e}}{c\,d^2\,e}&-\frac{c\,d}{e}&-\frac{e^2\sqrt{-\frac{c^6\,d^6}{e^2}}}{c\sqrt{c^4\,d^5\,e}} &d&e\\*[2mm]
-\frac{c\,d}{e}&c&\frac{\sqrt{c^4\,d^5\,e}}{c\,d^2\,e}&e&-\frac{e^2\sqrt{-\frac{c^6\,d^6}{e^2}}}{c\sqrt{c^4\,d^5\,e}} &d\\*[2mm]
\frac{1}{d}&-\frac{c\sqrt{c^4\,d^5\,e}}{e^2\sqrt{-\frac{c^6\,d^6}{e^2}}}&\frac{1}{e}&-\frac{c\,d^2\,e}{\sqrt{c^4\,d^5\,e}}&-\frac{1}{c}&\frac{e}{c\,d}\\*[2mm]
\frac{1}{e}&\frac{1}{d}& -\frac{c\sqrt{c^4\,d^5\,e}}{e^2\sqrt{-\frac{c^6\,d^6}{e^2}}} &\frac{e}{c\,d}&-\frac{c\,d^2\,e}{\sqrt{c^4\,d^5\,e}}&-\frac{1}{c}\\*[2mm]
- \frac{c\sqrt{c^4\,d^5\,e}}{e^2\sqrt{-\frac{c^6\,d^6}{e^2}}}  &\frac{1}{e}&\frac{1}{d}&-\frac{1}{c}&\frac{e}{c\,d}&-\frac{c\,d^2\,e}{\sqrt{c^4\,d^5\,e}}
\end{array}\right]\label{mcde2}\end{eqnarray}
and its reduced polynomial is
\begin{eqnarray}
pol(f_{+},\,a_{-})= y^3+\frac{\sqrt{3/2}(c^2d-e)\sqrt{c^4d^5e}}{c^3\,d^3\,e} y^2+\frac{(c^4d^2+e^2)y}{c^2\,d\,e} - \nonumber\\ 
\frac{c^2d(d^3-e^3)(c^2d+e)^3- 4(c^6d^3-e^3)\sqrt{c^4d^5e}}{6\,\sqrt{6\,}c^5\,d^4\,e^3}~~~\label{pol2}\end{eqnarray}
From the reduced equations (\ref{pol1}) and  (\ref{pol2}) one sees that the above two matrices are not unitary equivalent.

There are two more matrices generated by the pairs $(a_{+},\,f_{+})$ and $(a_{+},\,f_{-})$, and for them we write only their reduced polynomials
\begin{eqnarray}
pol(a_{+},\,f_{+})=  y^3+\frac{\sqrt{3/2}(c^2d-e)\sqrt{-c^4d^6}\sqrt{-\frac{c^6d^5}{e}}}{c^6d^6} y^2+\frac{(c^4 d^2+e^2)}{c^2\,d\,e} y -\nonumber\\ 
  \frac{c^5d^4(d^3-e^3)(c^2d+e)^3-4(c^6d^3-e^3)\sqrt{-c^4d^6}\sqrt{-\frac{c^6d^5}{e}}}{6\,\sqrt{6}\,c^8\,d^7\,e^3}~~~ \label{pol3}\end{eqnarray}

\begin{eqnarray}
pol(a_{+},\,f_{-})= y^3-\frac{\sqrt{3/2}(c^2d-e)\sqrt{-c^4d^6}\sqrt{-\frac{c^6d^5}{e}}}{c^6d^6} y^2+\frac{(c^4 d^2+e^2)}{c^2\,d\,e}y -\nonumber\\
\frac{c^5 d^4(d^3-e^3)(c^2d+e)^3 +4(c^6d^4-e^3)\sqrt{-c^4 d^6}\sqrt{-\frac{c^6d^5}{e}}}{6\,\sqrt{6}\,c^8\,d^7\,e^3}~~~\label{pol4}\end{eqnarray}

The reduced four polynomials, (\ref{pol1}) and (\ref{pol2})-(\ref{pol4}), are a strong argument that the four-parameter matrices have only four independent solutions for each pair $a_{\pm}$ and  $f_{\pm}$, and those similar to. The above three-parameter matrices obtained from four-parameter familes show that the number of these matrices is at least 60.

From the second factor of relation (\ref{hu}) one get similar results. For example the solutions $b = c\,e^2/d^2$ and $c= b\,d^2/e^2$ lead also to two matrices which are a litlle more complicated  than the   $D_{61}$ and  $D_{62}$ matrices. The other solutions $d=\pm \sqrt{c\,e^2/b}$ and  $e=\pm \sqrt{b\,d^2/c}$ could lead in principle to four matrices. Their spectral equations are also more complicated.

All these matrices are different from the other three-parameter matrices  found in literature, e.g., see paper \cite{K2}.
They  cannot be ``simplyfied'' to  simpler forms; in some cases the  analytic continuation can be done by ``hand'', but the spectra of the new matrices are in general different from the spectra of matrices one started with. From    the above matrices one get many matrices depending on two parameters by making equal two of them.

 As concerns the number of three-parameter matrices the second factor of relation (\ref{hu}) generates also four different matrices similar to matrices (\ref{mcde})-(\ref{pol4}) such that the total number of such matrices is at least 
$8\times 15 =120$.

Let see what happens when  the matrix $M_6$ is brought to its standard form. One gets
\begin{eqnarray}
M_{6s}=\left[\begin{array}{rrrrrr}
1&1&1&1&1&1\\*[2mm]
1&\frac{a^2}{bc}&\frac{ab}{c^2}&\frac{af}{cd}&\frac{ad}{ce}&\frac{ae}{cf}\\*[2mm]
1&\frac{ac}{b^2}&\frac{a^2}{bc}&\frac{ae}{bd}&\frac{af}{be}&\frac{ad}{bf}\\*[2mm]
1&\frac{ad}{bf}&\frac{ad}{ce}&-1&-\frac{ad}{ce}&-\frac{ad}{bf}\\*[2mm]
1&\frac{ae}{bd}&\frac{ae}{cf}&-\frac{ae}{bd}&-1&-\frac{ae}{cf}\\*[2mm]
1&\frac{af}{be}&\frac{af}{cd}&-\frac{af}{cd}&-\frac{af}{be}&-1\end{array}\label{hs}\right]\end{eqnarray}

The constraints satisfied by $M_{6s}$ matrix are the same as for the matrix $M_6$, i.e. the equations (\ref{con}) and (\ref{con1}), and after their fulfilment its spectral equations have to be in general different from those generated by the $M_6$ matrix. 

As an  example we generated  matrices by using the pairs $f_{\pm}$ and  $a_{\pm}$ into the matrix  $D_{6s}$, and we made two choices, $b = -cd/e$  and $b = ce^2/d^2$, to obtain  matrices similar to  matrices whose reduced polynomials are (\ref{pol1}) and (\ref{pol2})-(\ref{pol4}). The unexpected result was that there  is only {\em one} polynomial, and {\it one} eigenvalue set 
\begin{eqnarray}
ec(M_{6s})= x^6+2\sqrt{\frac{2}{3}} x^5+\frac{5}{3} x^4-\frac{5}{3} x^2-2\sqrt{\frac{2}{3}} x-1=0,~~Sp(M_{6s})=\left[-1,1,\left(-\frac{1+i \sqrt{5}}{\sqrt{6}}\right)^2,\left(\frac{-1+i \sqrt{5}}{\sqrt{6}}\right)^2\right] \label{stif}
\end{eqnarray}
which does not depend on $f_{\pm}$ and  $a_{\pm}$, showing that  the dephased  and its undephased forms  are not unitary equivalent.

An explanation could be the following. The coefficient of the $x^3$ power in the spectral polynomial of the matrix which depends on the four free parameters, $b,\,c,\,d,\,e$, has as factor the relation (\ref{hu}) multliplied by a huge expression. This explains the absence of the   $x^3$ power in equation (\ref{stif}), but does no explain the other constant terms.

\newpage

\section{Higher dimensional Hadamard matrices}

For $n= 8$  the analog of matrix (\ref{m6}) is the following

\begin{eqnarray}
M_8=\left[\begin{array}{rrrrrrrr}
a&b&c&d&e&f&g&h\\*[2mm]
d&a&b&c&h&e&f&g\\*[2mm]
c&d&a&b&g&h&e&f\\*[2mm]
b&c&d&a&f&g&h&e\\*[2mm]
\frac{1}{e}&\frac{1}{h}&\frac{1}{g}&\frac{1}{f}&-\frac{1}{a}&-\frac{1}{d}&-\frac{1}{c}&-\frac{1}{b}\\*[2mm]
\frac{1}{f}&\frac{1}{e}&\frac{1}{h}&\frac{1}{g}&-\frac{1}{b}&-\frac{1}{a}&-\frac{1}{d}&-\frac{1}{c}\\*[2mm]
\frac{1}{g}&\frac{1}{f}&\frac{1}{e}&\frac{1}{h}&-\frac{1}{c}&-\frac{1}{b}&-\frac{1}{a}&-\frac{1}{d}\\*[2mm]
\frac{1}{h}&\frac{1}{g}&\frac{1}{f}&\frac{1}{e}&-\frac{1}{d}&-\frac{1}{c}&-\frac{1}{b}&-\frac{1}{a}\end{array}\right]\label{m8}\end{eqnarray}
and relation (\ref{inv2}) leads to three constraints
\begin{eqnarray}
&& a b c d e^2f g +a^2 b c e f g h+b^2 c d e f g h+a c^2d e f g h+a b d^2e f g h + a b c d f^2g h+a  b c d e g^2h+a b c d e f h^2=0\label{hh1}\\
&& a b c d e f^2g+a b c d e^2f h+a b^2c e f g h+a^2b d e f g h+b c^2d e f g h+a c d^2e f g h+a b c d f g^2h+a b c d e g h^2=0\label{e2}\\
&&a b c d e f g^2+a b c d e f^2 h+a b c d e^2g h+a b c^2e f g h+a b^2d e f g h+a^2c d e f g h+b c d^2e f g h+a b c d f g h^2=0\label{a3}
\end{eqnarray}

Solving equation (\ref{hh1}) with respect to $h$ parameter one get two  solutions. When these solutions are introduced in equation   (\ref{a3}) one gets one equation which does not depend on the square root, similar to the case $n=6$
\begin{eqnarray}
&&(a^2bcef+b^2cdef+ac^2def+abd^2ef+abcdf^2+abcdeg)g^2-\nonumber\\
&&(abcdf^2+abcdeg+abc^2fg+ab^2dfg+a^2cdfg+bcd^2fg)e^2 \label{e31}\end{eqnarray}

If the same solution is introduced into equation (\ref{e2}) one gets an equation that depends on the square root entering the $h$ parameter. By multiplying it with its conjugate one gets a huge polynomial, $P$, that depends on powers, $a^4, \, b^4,\,\dots,g^4$, in the corresponding monomials. If one solution of the equation (\ref{e31}) is introduced into the above $P$ polynomial one finds an equation  that depends on square root and the third power of the square root expressions. Even if  such an equation could be transformed into a polynomial, each power of the remaining parameters will be greater than eight, such that one cannot solve analytically such an equation. 

On could say that the form of (\ref{m8}) matrix is not  appropiate for getting Hadamard matrices; we tried a few other forms but the obtained constraints are at least four, such that in the best case the resulting matrix will depend on the same number of parameters as matrices obtained from (\ref{m6}).

Thus in this case we have to  make use of  the alternative  formula, see \cite{D}, equation (8), of the form

\begin{eqnarray}
M_1=\left[ \begin{array}{cc}
A&~D\,B\\
A&-D\,B\end{array}\right]\label{di}\end{eqnarray}
where $A$ and $B$ are Hadamard matrices, and $D$ is a $n$-dimensional diagonal matrix containing phases.

By using the results from section {\bf 2}  we can take two 4-dimensional matrices
\begin{eqnarray}\begin{array}{cc}
A_4=\left[\begin{array}{rrrr}
\frac{b d}{c}&b&c&d\\*[2mm]
-b&\frac{b d}{c}&-d&c\\*[2mm]
\frac{1}{c}&-\frac{1}{d}&-\frac{c}{b d}&\frac{1}{b}\\*[2mm]
\frac{1}{d}&\frac{1}{c}&-\frac{1}{b}&-\frac{c}{b d}\end{array}\right]\,,&
B_4=\left[\begin{array}{rrrr}
f&g&h&\frac{fh}{g}\\*[2mm]
-g&f&-\frac{fh}{g}&h\\*[2mm]
\frac{1}{h}&-\frac{g}{f h}&-\frac{1}{f}&\frac{1}{g}\\*[2mm]
\frac{g}{f h}&\frac{1}{h}&-\frac{1}{g}&-\frac{1}{f}\end{array}\right]\end{array}
\end{eqnarray}
and construct with them an eight-dimensional matrix by using relation (\ref{di}). It has the form

\begin{eqnarray}
D_{8a}(b,c,d,f,g,h)=\left[\begin{array}{rrrrrrrr}
\frac{b d}{c}&b&c&d&f&g&h&\frac{fh}{g}\\*[2mm]
-b&\frac{b d}{c}&-d&c&-g&f&-\frac{fh}{g}&h\\*[2mm]
\frac{1}{c}&-\frac{1}{d}&-\frac{c}{b d}&\frac{1}{b}&\frac{1}{h}&-\frac{g}{f h}&-\frac{1}{f}&\frac{1}{g}\\*[2mm]
\frac{1}{d}&\frac{1}{c}&-\frac{1}{b}&-\frac{c}{b d}&\frac{g}{f h}&\frac{1}{h}&-\frac{1}{g}&-\frac{1}{f}\\*[2mm]
\frac{b d}{c}&b&c&d&-f&-g&-h&-\frac{fh}{g}\\*[2mm]
-b&\frac{b d}{c}&-d&c&g&-f&\frac{fh}{g}&-h\\*[2mm]
\frac{1}{c}&-\frac{1}{d}&-\frac{c}{b d}&\frac{1}{b}&-\frac{1}{h}&\frac{g}{f h}&\frac{1}{f}&-\frac{1}{g}\\*[2mm]
\frac{1}{d}&\frac{1}{c}&-\frac{1}{b}&-\frac{c}{b d}&-\frac{g}{f h}&-\frac{1}{h}&\frac{1}{g}&\frac{1}{f}\end{array}\label{h8a}\right]\end{eqnarray}

The above matrix depends on 6 free parameters and if the $B$-matrix is multiplied by $D$ the new matrix depends on 10  parameters. 

The spectral equation of the matrix (\ref{h8a}) can be written but it is too long, so we chose some particular values for all the six parameters, $b=1,\,c=i\,,d=-i,\,f=\frac{1+i}{\sqrt{2}},\,g=\frac{1-i}{\sqrt{2}},\,h=-1$, which generate the matrix
\begin{eqnarray}
D_{81}(1,i,-i,e^{\frac{\pi i}{4}},e^{-\frac{\pi i}{4}} ,-1)=\left[\begin{array}{rrrrrrrr}
-1&1&i&-i&\frac{1+i}{\sqrt{2}}&\frac{1-i}{\sqrt{2}}&-1&-i\\*[2mm]
-1&-1&i&i&-\frac{1-i}{\sqrt{2}}&\frac{1+i}{\sqrt{2}}&i&-1\\*[2mm]
-1&-i&1&1&-1&-i&-\frac{1-i}{\sqrt{2}}&\frac{1+i}{\sqrt{2}}\\*[2mm]
i&-i&-1&1&i&-1&-\frac{1+i}{\sqrt{2}}&-\frac{1-i}{\sqrt{2}}\\*[2mm]
-1&1&i&-i&-\frac{1+i}{\sqrt{2}}&-\frac{1-i}{\sqrt{2}}&1&i\\*[2mm]
-1&-1&i&i&\frac{1-i}{\sqrt{2}}&-\frac{1+i}{\sqrt{2}}&-i&1\\*[2mm]
-i&-i&1&1&1&i&\frac{1-i}{\sqrt{2}}&-\frac{1+i}{\sqrt{2}}\\*[2mm]
i&-i&-1&1&-i&1&\frac{1+i}{\sqrt{2}}&\frac{1-i}{\sqrt{2}}
\end{array}\label{h8b}\right]\end{eqnarray}
The reduced equation has the form
\begin{eqnarray}
y^4-i y^3+\frac{4+\sqrt{2}}{2}-i\frac{6+\sqrt{2}}{4}+\frac{1+\sqrt{2}}{2}=0\label{nh8}\end{eqnarray}
whose solutions are 
\begin{eqnarray}Sol(ec(\ref{nh8}))=\left[y_1 = -i \sqrt{2},\,y_2 = i\frac{2+\sqrt{2}}{2},\,y_3 = i\frac{(\sqrt{2}+\sqrt{10})}{4},\,y_4 = i\frac{(\sqrt{2}-\sqrt{10})}{4}\right]\label{sol}\end{eqnarray}
and the spectrum of matrix (\ref{h8b}) is given by

\begin{eqnarray}
Sp(D_{81}(1,i,-i,e^{\frac{\pi i}{4}},e^{-\frac{\pi i}{4}} ,-1)~~=~~\left[x_{1,2}=\frac{i\pm1}{\sqrt{2}},\,x_{3,4}=
-\frac{i(2+\sqrt{2})\pm \sqrt{10-4\sqrt{2}}}{4},\,\right.\nonumber\\
\left.x_{5,6}=\frac{\pm 2\sqrt{13-\sqrt{5}}-i\sqrt{2}(1+\sqrt{5})}{8},\,x_{7,8}= \frac{i(\sqrt{10}-\sqrt{2})\pm 2 \sqrt{13+\sqrt{5}}}{8}\right]\end{eqnarray}

One problem is to find how many 8-dimensional  matrices, similar to (\ref{h8a}) can be found. In section {\bf 2} we have shown that the relation (\ref{ort}) has eight solutions. Thus all these matrices can be used as matrix $A$, and by changing the parameters as in the above matrix $B$ one arrives at 64 matrices similar to (\ref{h8a}). 

The relation (\ref{di}) can be used again with the matrix (\ref{h8a}) to obtain matrices of dimensions $ H_{2^n 8}$  for $n = 1,\,2,\,\dots$; for example the matrix $H_{16}$ will depend on 10+10+8=28 free parameters, etc.

The same thing can be done with the 6-dimensional Hadamard matrices. All the 4-parameter six-dimensional matrices can be used to obtain 12-dimensional Hadamard matrices which depend on fourteen parameters. These matrices are also large and  cannot be written down, but numerical matrices can be. The spectral equation will be 12-dimensional and its reduced equation is  6-dimensional, so one cannot find for it  analytic solutions.
A numerical example is the following
\begin{eqnarray}\begin{array}{cc}
A_{6} = \left[\begin{array}{cccccc}
-\frac{1-i}{\sqrt{2}}&1&i&-i&-1&\frac{1-i}{\sqrt{2}}\\*[2mm]
i&-\frac{1-i}{\sqrt{2}}&1&\frac{1-i}{\sqrt{2}}&-i&-1\\*[2mm]
1&i&-\frac{1-i}{\sqrt{2}}&-1&\frac{1-i}{\sqrt{2}}&-i\\*[2mm]
i&\frac{1+i}{\sqrt{2}}&-1&\frac{1+i}{\sqrt{2}}&i&-1\\*[2mm]
-1&i&\frac{1+i}{\sqrt{2}}&-1&\frac{1+i}{\sqrt{2}}&i\\*[2mm]
\frac{1+i}{\sqrt{2}}&-1&i&i&-1&\frac{1+i}{\sqrt{2}}\end{array}\right]\,,&
B_6= \left[\begin{array}{cccccc}
1&\frac{1+i}{\sqrt{2}}&\frac{i\sqrt{1+i}}{2^{1/4}}&\frac{2^{1/4}}{\sqrt{1+i}}&\frac{1-i}{\sqrt{2}}&-1\\*[2mm]
\frac{i\sqrt{1+i}}{2^{1/4}} &1&\frac{1+i}{\sqrt{2}}&-1&\frac{2^{1/4}}{\sqrt{1+i}}&\frac{1-i}{\sqrt{2}}\\*[2mm]
\frac{1+i}{\sqrt{2}}&\frac{i\sqrt{1+i}}{2^{1/4}}&1&\frac{1-i}{\sqrt{2}}&-1&\frac{2^{1/4}}{\sqrt{1+i}}\\*[2mm]
\frac{\sqrt{1+i}}{2^{1/4}}&-1&\frac{1+i}{\sqrt{2}}&-1&\frac{(1+i)^{3/2}}{2^{3/4}}&-\frac{1-i}{\sqrt{2}}\\*[2mm]
\frac{1+i}{\sqrt{2}}&\frac{\sqrt{1+i}}{2^{1/4}}&-1&-\frac{1-i}{\sqrt{2}}&-1&\frac{(1+i)^{3/2}}{2^{3/4}}\\*[2mm]
-1&\frac{1+i}{\sqrt{2}}&\frac{\sqrt{1+i}}{2^{1/4}}&\frac{(1+i)^{3/2}}{2^{3/4}}&-\frac{1-i}{\sqrt{2}}&-1
\end{array}\right]\label{ab6}\end{array}
\end{eqnarray}

The spectrum of the corresponding numerical $D_{12}$ matrix generated by $A_{6}$ and  $B_{6}$ can be found  numerically by using, e.g. Mathematica, and in this numerical case the spectrum  is simple. As in the preceding $n= 8$ case, for $n\ge 12$ one can use relation (\ref{di}) to obtain Hadamard matrices with dimensions
$H_{2^n 12}$ for any integer $n\ge 1$.

\section{Conclusion}

Starting with a carefully chosen form of the 6-dimensional matrix, (\ref{m6}), we got many  6-dimensional Hadamard matrices that depend on four parameters. The peculiarity of all of them is that they are long matrices which can be stored and viewed only in an electronic format. In fact after the substitution of two chosen parameters, e.g. $b$ and $f$, the complex Hadamard matrix $M_6= M_6(a,c,d,e) $ depends on four independent parameters. However many of its entries become so long, that from a typographical point of view $M_6$ can hardly be presented as a matrix, but it can be used in an algebraic computer program.

 The number of four- and three-parameter matrices being quite big there is a hope that the unsolved problem of mutually unbiased bases could be solved for six-level systems. This problem received a special attention of many groups, but despite substantial efforts until now no relevant breakthrough was provided, although papers appear continuosly, the last being \cite{JMM} and \cite{BB}.

 Looking at the form of $M_{6}$ matrix one sees that by our procedure one cannot obtain all the the 6-dimensional matrices that depend on four parameters, one of them is the Agaian matrix, see \cite{A}, page 112. Thus other approaches are necessary to find  all those matrices which include  Agaian matrix.

\begin{acknowledgments}
It is a pleasure to thank K.  $\dot{\rm Z}$yczkowski for a critical reading of the manuscript and for a few judiciuos suggestions, that lead to the improvement of the paper. We thank  I. Bengtsson who informed me that the form of $M_6$ matrix (\ref{m6}) appears also in paper \cite{Sz}.

We acknowledge partial  support from Project PN09370102/2009 and Contract Idei 121/2011
\end{acknowledgments}

\end{document}